\documentclass{amsart}

\usepackage{amssymb,epsf}

\newcommand{\be}{\begin{equation}}
\newcommand{\ee}{\end{equation}}
\newcommand{\bea}{\begin{eqnarray}}
\newcommand{\eea}{\end{eqnarray}}
\newcommand{\beas}{\begin{eqnarray*}}
\newcommand{\eeas}{\end{eqnarray*}}

\newtheorem{thm}{Theorem}

\newtheorem{prop}[thm]{Proposition}
\newtheorem{cor}[thm]{Corollary}

\title{Not so non-renormalizable gravity${}^*$}
\author{Dirk Kreimer}
\address{kreimer@ihes.fr, IHES, 35 rte. de Chartres, 91440 Bures-sur-Yvette, France (http://\ www.ihes.fr)
and Boston U.\ (http://math.bu.edu)}
\thanks{${}^*$Work supported in parts by grant NSF-DMS/0603781. Author supported by CNRS}
\begin{document}
\maketitle
\begin{abstract}
We review recent progress with the understanding of quantum fields, including ideas \cite{AnnQG} how gravity might turn out to be a renormalizable theory after all.
\end{abstract}
\section{Introduction}
Renormalizable perturbative quantum field theories are embarrassingly successful in describing observed physics.
Whilst their mathematical structure is still a challenge albeit an entertaining one, they are testimony to some of the finest achievements in our understanding of nature. The physical law as far at is insensitive to the surrounding geometry seems completely described by such theories.
Alas, if we incorporate gravity, and want to quantize it, we seem at a loss.

In this talk, we report on some recent work \cite{AnnQG} which might give hope. Our main purpose is to review the basic idea and to put it into context.

As in \cite{AnnQG}, we will proceed by a comparison of the structure of a renormalizable theory, quantum electrodynamics in four dimensions, and gravity.

It is the role of the  Hochschild cohomology \cite{BergbK} in those two different situations which leads to surprising new
insights. We will discuss them
at an elementary level for the situation of pure gravity. We also allow, in the spirit of the workshops where this material was presented, for the freedom to muse about conceptual consequences at the end.
\subsection*{Acknowledgments}
This short contribution is based on talks given in Leipzig ({\it Recent Developments in QFT}, MPI Math.\ in the Sciences, Leipzig, July 20-22 2007, B.\ Fauser, J.\ Tolksdorf, E.\ Zeidler, eds.) and Bonn ({\it Conference on Combinatorics and Physics}, MPI Math., Bonn, March 19-23 2007, K.\ Ebrahimi-Fard, M.\ Marcolli, W.\ van Suijlekom, eds.). Its a pleasure to thank all the organizers for their efforts and hospitality.
\section{The structure of Dyson--Schwinger Equations in QED${}_4$}
\subsection{The Green functions}
Quantum electrodynamics in four dimensions of space-time (QED${}_4$) is
described in its short-distance behaviour by four Green functions \be G^{\bar{\psi}\gamma\cdot\partial\psi},G^{m\bar{\psi}\psi},G^{\bar{\psi}\gamma\cdot A\psi},G^{\frac{1}{4}F^2},\ee
corresponding to the four monomials
in its Lagrangian \be L=\bar{\psi}\gamma\cdot\partial\psi-\bar{\psi}m\psi-\bar{\psi}\gamma\cdot A\psi-\frac{1}{4}F^2.\ee
Here, $G^i=G^i(\alpha,L)$, with $\alpha$ the fine structure constant and $L=\ln q^2/\mu^2$. We hence work in a MOM scheme, subtract at $q^2=\mu^2$,
project the vertex function to its scalar formfactor $G^{\bar{\psi}\gamma\cdot\partial\psi}$ with UV divergences evaluated at zero photon momentum.
Similarly the other Green functions are normalized as to be the multiplicative quantum corrections to the tree level monomials above, in momentum space.

In perturbation theory, the degree of divergence of a graph $\Gamma$ with $f$ external
fermion lines and $m$ external photon lines in $D$ dimensions is \be
\omega_D(\Gamma)=\frac{3}{2}f+m-D-(D-4)(|\Gamma|-1)\Rightarrow
\omega_4(\Gamma)=\frac{3}{2}f+m-4.\ee This is independent of the
loop number for QED${}_4$, $D=4$, and is a sole function of the
number and type of external legs. $\omega_D(\Gamma)$ determines the number of derivatives with respect to masses or external momenta needed to render
a graph logarithmically divergent, and hence identifies the top-level residues which drive the iteration of Feynman integrals according to the quantum equations of motion \cite{KY2}.

We define these four Green functions as an evaluation by renormalized Feynman rules of a series of one-particle irreducible (1PI)
Feynman graphs $\Gamma\in {\mathcal{FG}}_i$.  These series are determined as a fixpoint of the following system in Hochschild cohomology.
\bea
X^{\bar{\psi}\gamma\cdot\partial\psi} & = & 1-\sum_{k=1}^\infty\alpha^k B_+^{\bar{\psi}\gamma\cdot\partial\psi,k}(X^{\bar{\psi}\gamma\cdot\partial\psi}Q^{2k}(\alpha)),\\
X^{\bar{\psi}\gamma\cdot A\psi} & = & 1+\sum_{k=1}^\infty\alpha^k B_+^{\bar{\psi}\gamma\cdot A\psi,k}(X^{\bar{\psi}\gamma\cdot A\psi}Q^{2k}(\alpha)),\\
X^{\bar{\psi}m\psi} & = & 1-\sum_{k=1}^\infty\alpha^k B_+^{\bar{\psi}m\psi,k}(X^{\bar{\psi}m\psi}Q^{2k}(\alpha)),\\
X^{\frac{1}{4}F^2} & = & 1-\sum_{k=1}^\infty\alpha^k B_+^{\frac{1}{4}F^2,k}(X^{\frac{1}{4}F^2}Q^{2k}(\alpha)).
\eea
Here, \be B_+^{i,k} = \sum_{|\gamma|=k,  \Delta^\prime(\gamma)=0,\gamma\in {\mathcal{FG}}_i} B_+^\gamma,\;\forall i\in {\mathcal R}_{\textrm{QED}}, \ee
a sum over all Hopf algebra primitive graphs with given loop number $k$ and contributing to superficially divergent amplitude $i$,
and
\be B_+^\gamma(X)=\sum_{\Gamma\in <\Gamma>}\frac{{\textbf{bij}(\gamma,X,\Gamma)}}{|X|_\vee}\frac{1}{\textrm{maxf}(\Gamma)}\frac{1}{(\gamma|X)}\Gamma,\label{def}\ee
where maxf$(\Gamma)$ is the number of maximal forests of $\Gamma$, $|X|_\gamma$ is the number of distinct graphs obtainable by permuting edges of $X$, $\textrm{bij}(\gamma,X,\Gamma)$ is the number of bijections of external edges of $X$ with an insertion place in $\gamma$ such that the result is $\Gamma$,
and finally $(\gamma|X)$ is the number of insertion places for $X$ in $\gamma$ \cite{anatomy}, and
 \be {\mathcal R}_{\textrm{QED}}=\{\bar{\psi}\gamma\cdot\partial\psi,\bar{\psi}\gamma\cdot A\psi,m\bar{\psi}\psi,\frac{1}{4}F^2\}.\ee

Also, we let \be Q=\frac{X^{\bar{\psi}\gamma\cdot A\psi}}{ X^{\bar{\psi}\gamma\cdot\partial\psi}\sqrt{X^{\frac{1}{4}F^2}}}.\ee

The resulting maps $B_+^{i,K}$ are Hochschild closed \be bB_+^{i,K}=0,\label{closed}\ee in the sense of \cite{tor}.
We have in fact \be \Delta(B_+^\gamma(X))=\sum_\Gamma n_{\Gamma,X,\gamma} \Gamma\ee
where $n_{\Gamma,X,\gamma}$ can be  determined from (\ref{def},\ref{closed}).

Furthermore, one can choose a basis of primitives $\gamma$ \cite{KY2} such that their Mellin transforms $M_\gamma(\rho)$ have the form
\be
M_\gamma(\rho)=\int \iota_\gamma(k_i;q)_{|_{q^2=\mu^2}} \prod_{s=1}^{|\gamma|}\frac{[k_s^2/\mu^2]^{-\rho/|\gamma|}d^4k_i}{(2\pi)^4}\;\textrm{for}\;1>\Re(\rho)>0,\ee
where the integrand $\iota_\gamma$ is a function of internal momenta $k_i$ and an external momentum $q$, subtracted at $q^2=\mu^2$.

The Dyson Schwinger equations then take the form
\be
G_R^i(\alpha,L)=1\pm\lim_{\rho\to 0}\left[\sum_k\alpha^k \sum_{|\gamma|=k} G_r^i(\alpha,\partial_\rho){\mathcal Q}(\alpha,\partial_\rho)M_\gamma(\rho)\left[\left(\frac{q^2}{\mu^2}\right)^{-\rho}-1\right]\right],\ee
where \be \Phi_R(X^i)=G_R^i(\alpha,L),\ee
and
\be  \Phi_R(Q)={\mathcal Q}(\alpha,L),\ee
is the invariant charge, all calculated with renormalized Feynman rules in the MOM scheme.
Note that such subtractions on skeleton kernels not only provide a means to investigate non-perturbative aspects of Green functions \cite{KY2},
but are also well-founded mathematically \cite{BlKr}.

\subsection{Gauge theoretic aspects}

Using Ward identities, we can reduce the set ${\mathcal R}_{\textrm{QED}}=\{\bar{\psi}\gamma\cdot\partial\psi,m\bar{\psi}\psi,\bar{\psi}\gamma\cdot A\psi,\frac{1}{4}F^2\}$
to three elements upon identifying $G^{\bar{\psi}\gamma\cdot\partial\psi}=G^{\bar{\psi}\gamma\cdot A\psi}$.
Using the Baker--Johnson--Willey gauge \cite{BJW1} we can furthermore trivialize
\be
G^{\bar{\psi}\gamma\cdot\partial \psi}=G^{\bar{\psi}\gamma\cdot A\psi}=1.
\ee
Using their work again \cite{BJW2}, we have that $m\bar{\psi}\psi$ can be ignored in ${\mathcal R}_{\textrm{QED}}$.

We are hence left with the determination of a single gauge-independent Green function $G^{\frac{1}{4}F^2}$ which in the MOM scheme takes the form
\be G^{\frac{1}{4}F^2}(\alpha,L)=1-\sum_{k=1}^\infty \gamma_k(\alpha)L^k,\ee
and the renormalization group determines \cite{KY1}
\be \gamma_k(\alpha)=\frac{1}{k}\gamma_1(\alpha)(1-\alpha\partial_\alpha)\gamma_{k-1}(\alpha).\ee
Here, $\gamma_1(\alpha)=2\psi(\alpha)/\alpha$, where $\psi(\alpha)$ is the MOM scheme $\beta$-function of QED,
which is indeed half of the anomalous dimension $\gamma_1$ of the photon field in that scheme.

One can show that $\gamma_1(\alpha)$ as a perturbative series ($\gamma_1(\alpha)=\sum_{j=1}^\infty \gamma_{1,j}\alpha^j$) is Gevrey--1 and
that the series  $\sum_{j=1}^\infty \gamma_{1,j}\alpha^j/j!$ has a finite radius of convergence, with a bound involving the lowest order contribution
of the $\beta$-function and the one-instanton action \cite{KY2}.

Furthermore, $\gamma_1(\alpha)$ fulfills \cite{KY2}
\be \gamma_1(\alpha)=P(\alpha)-\gamma_1(\alpha)(1-\alpha\partial_\alpha)\gamma_1(\alpha),\ee
an equation which has been studied in detail recently \cite{KY3}, with more of its analytic structure to be exhibited there.
In particular, the presence or absence of Landau poles beyond perturbation theory was clarified in terms of concrete conditions on the asymptotics of
$P(\alpha)$ \cite{KY3}.
In this equation, $P(\alpha)$ is obtained from the primitives of the Hopf algebra
\be P(\alpha)=\sum_\gamma \alpha^{|\gamma|}\lim_{\rho\to 0}\rho M_\gamma(\rho),\ee
 and $P(\alpha)$ is known perturbatively as a fifth order polynomial \cite{Surg} and its asymptotics have been conjectured long ago \cite{Zetal}.

This finishes our summary of QED$_{4}$ as a typical renormalizable theory.
\subsection{Non-abelian gauge theory}
 The above approach to Green functions remains valid for a non-abelian gauge theory with the definition of a single invariant charge
 ${\mathcal Q}(\alpha,L)$ being the crucial requirement. This can be consistently  done, \cite{anatomy}, upon recognizing that the celebrated Slavnov--Taylor
 identities for the couplings fulfill
 \be \frac{S_R^\phi(X^{\bar{\psi}\gamma\cdot A\psi})}{S_R^\phi(X^{\bar{\psi}\gamma\cdot\partial\psi})}=\frac{S_R^\phi(X^{AA\partial A})}{S_R^\phi(X^{\partial A\partial A})}
 =\frac{S_R^\phi(X^{AAAA})}{S_R^\phi(X^{AA\partial A})}=\frac{S_R^\phi(X^{\bar{\phi}A\cdot\partial\phi})}{S_R^\phi(X^{\bar{\phi}\Box\phi})},\ee
 for the set of amplitudes \be {\mathcal{R}}_{\textrm{QCD}}=\{DADA,\bar{\psi}\gamma\cdot\partial\psi,\bar{\phi}\Box\phi,AADA,AAAA,\bar{\phi}A\cdot\partial\phi,\bar{\psi}\gamma\cdot A\psi\},\ee
 needing renormalization in QCD.

This allows to define a Hochschild cohomology on the sum of graphs at a given loop order, and hence to obtain multiplicative renormalization
 in this language from the resulting coideals in the Hopf algebra \cite{anatomy,vSlk}.

Note that the structure of the sub-Hopf algebras underlying this approach \cite{tor,KY1} implies that the elements $X^i(\alpha)$ close under the coproduct. A general classification of related sub-Hopf algebras has been recently obtained by Loic Foissy \cite{Foissy}. He considers only the case that the lowest order Hochschild cocycle is present in the combinatorial Dyson--Schwinger equations, but his study is rather complete when augmented by the results of \cite{BergbK}. Indeed, this allows to incorporate the next-to-leading order cocycles without any change to the structure of the theory.

\section{Gravity}
We consider pure gravity understood as a theory based on a graviton propagator and $n$-graviton couplings as vertices.
A fuller discussion incorporating ghosts and matter fields is referred to future work.
\subsection{Summary of results of \cite{AnnQG}}
\begin{cor} Let $|\Gamma|=k$. Then $\omega(\Gamma)=-2(|\Gamma|+1)$.\end{cor}
This is a significant change from the behavior of a renormalizable theory: in the renormalizable case, each graph contributing to the same amplitude $i$ has the same powercounting degree regardless of the loop number. Here, we have the dual situation: the loop number determines the powercounting degree, regardless of the amplitude.
\begin{thm} The set $d_{\omega(\Gamma)}$ contains no primitive element beyond one loop.\end{thm}
The set $d_{\omega(\Gamma)}$ is determined as a set of dotted graphs, with dots representing $\omega(\Gamma)$ derivatives with respect to masses or external momenta such that the corresponding integrand $\iota_\Gamma$ is overall log-divergent. Whilst in a renormalizable theory, we find for each
amplitude in the finite set ${\mathcal R}$ primitives at each loop order in $d_{\omega(\Gamma)}$, here we have an infinite set ${\mathcal R}$,
but only one primitive in it.
\begin{prop}
The relations \be \frac{X^{n+1}}{X^n}=\frac{X^n}{X^{n-1}},\;n\geq 3,\ee
define a sub-Hopf algebra with Hochschild closed one-cocycles
$B_+^{1,n}$.
\end{prop}
Here, $X^n$ is the sum of all graphs with $n$ external graviton lines. One indeed finds that the combinatorial Dyson--Schwinger equations for gravity provide a sub-Hopf algebra upon requiring these relations, in straightforward generalization of the situation in a non-abelian gauge theory.
\subsection{Comments}
\subsubsection{Gauss-Bonnet} The Gauss--Bonnet theorem ensures here, in the form
\be 0=\int_{\mathbb{M}}\sqrt{g}\left( R_{\mu\nu\rho\sigma}R^{\mu\nu\rho\sigma}-4R_{\alpha\beta}R^{\alpha\beta}+R^2\right)\ee
the vanishing
of the one-loop renormalization constants.
This does not imply the vanishing of the two-loop renormalization constants as their one-loop subdivergences are off-shell.
But it implies that the two-loop counter term has only a first order pole by the scattering type formula, in agreement with the vanishing of $\phi_{\textrm{off-shell}}(\gamma)\phi_{\textrm{on-shell}}(\Gamma/\gamma)$. Here, $\gamma$, $\Gamma/\gamma$ is the decomposition of $\Gamma$ into  one-loop graphs and $\phi_{\textrm{on/off-shell}}$ denotes suitable Feynman rules.

\subsubsection{Two-loop counterterm} Also, the universality of the two-loop counterterm suggests that indeed  \be\frac{Z^{\textrm{gr}_{n+1}}}{Z^{\textrm{gr}_{n}}}=
\frac{Z^{\textrm{gr}_{n}}}{Z^{\textrm{gr}_{n-1
}}},\;\textrm{with $Z^{\textrm{gr}_n}=S_R^\phi(X^n)$},\ee
holds for off-shell counterterms.
In particular, if we compute in a space of constant curvature and conformally reduced gravity which maintains many striking features of asymptotic safe gravity \cite{Martin,Martin2}, the above identities should hold for suitably defined characters: indeed, in such circumstances we can renormalize
using a graviton propagator which is effectively massive with the mass $\sqrt{R/6}$ provided by the constant curvature $R$,
and hence can renormalize at zero external momentum. Using the KLT relations \cite{bern}, this reduces the above identities to a (cumbersome) combinatorial exercise on one-loop graphs to be worked out in the future.

 Continuing this line of thought one expects that a single quantity, the $\beta$ function of gravity, exhibits short-distance singularities. If this expectation bears out, it certainly is in nice conceptually agreement with the expectation that in theories where gravity has a vanishing $\beta$ function, gravity is indeed a finite theory \cite{Dixon}.
\subsubsection{Other instances of gravity powercounting}
The appearance of Feynman rules such that the powercounting of vertex amplitudes  in ${\mathcal R}_{\textrm{V}}$ cancels the powercounting of
 propagator amplitudes in  ${\mathcal R}_{\textrm{E}}$, ${\mathcal R}={\mathcal R}_{\textrm{V}}\cup {\mathcal R}_{\textrm{E}}$, is not restricted to gravity. It indeed appears for example also in the field theoretic description of  Brownian fluids and glass possibly, which were recently described  at tree-level as a field theory \cite{VCCK}. The dynamics beyond tree level will involve renormalization with powercounting properties similar to the present discussion.

\end{document}